\documentclass[printer]{aa}  
\usepackage{graphicx}
\usepackage{txfonts}
\usepackage{natbib,twoopt}
\usepackage[breaklinks=false]{hyperref} 
\bibpunct{(}{)}{;}{a}{}{,}             
\usepackage{color}

\begin{document} 

\title{The AKARI IRC asteroid flux catalogue: \\ updated diameters and albedos}

\author{V. Al\'i-Lagoa\inst{1}
  \and
  T.~G. M{\"u}ller\inst{1}
  \and
  F. Usui\inst{2}
  \and
  S. Hasegawa\inst{3}
}

\institute{
  Max-Planck-Institut f{\"u}r extraterrestrische Physik, Giessenbachstrasse 1,
  85748 Garching, Germany\\
  \email{vali@mpe.mpg.de}
  \and
  Center for Planetary Science, Graduate School of Science, Kobe University,
  7-1-48 Minatojima-Minamimachi, Chuo-Ku, Kobe 650-0047, Japan
  \and
  Institute of Space and Astronautical Science, Japan Aerospace Exploration
  Agency, 3-1-1 Yoshinodai, Chuo-ku, Sagamihara 252-5210, Japan
}

\date{Accepted for publication in Astronomy \& Astrophysics (Dec. 2017)}


\abstract{  The AKARI IRC All-sky survey provided more than twenty thousand
  thermal infrared observations of over five thousand asteroids. Diameters and
  albedos were obtained by fitting an empirically calibrated version of the
  standard thermal model to these data. After the publication of the flux
  catalogue in October 2016, our aim here is to present the AKARI IRC all-sky
  survey data and discuss valuable scientific applications in the field of
  small body physical properties studies. As an example, we update the
  catalogue of asteroid diameters and albedos based on AKARI using the
  near-Earth asteroid thermal model (NEATM). We fit the NEATM to derive
  asteroid diameters and, whenever possible, infrared beaming parameters. We
  fit groups of observations taken for the same object at different epochs of
  the survey separately, so we compute more than one diameter for approximately
  half of
  the catalogue. We obtained a total of 8097 diameters and albedos for 5170
  asteroids, and we fitted the beaming parameter for almost two thousand of
  them. When it was not possible to fit the beaming parameter, we used a
  straight line fit to our sample's beaming parameter-versus-phase angle plot
  to set the default value for each fit individually instead of using a single
  average value. Our diameters agree with stellar-occultation-based diameters
  well within the accuracy expected for the model. They also match the previous
  AKARI-based catalogue at phase angles lower than 50 degrees, but we find a
  systematic deviation at higher phase angles, at which near-Earth and
  Mars-crossing asteroids were observed. The AKARI IRC All-sky survey is an
  essential source of information about asteroids, especially the large ones,
  since, it provides observations at different observation geometries,
  rotational coverages and aspect angles. For example, by comparing in
    more detail a few asteroids for which dimensions were derived from
  occultations, we discuss how the multiple observations per object may
  already provide three-dimensional information about elongated objects even
  based on an idealised model like the NEATM. Finally, we enumerate
  additional expected applications for more complex models, especially in
  combination with other catalogues. 
}

\keywords{minor planets, asteroids: general --
  surveys --
  infrared: planetary systems
}
\titlerunning{AKARI IRC: updated diameters and albedos}

\maketitle
%
\section{Introduction}\label{sec:intro}

Many asteroids are relatively intact fossils of the formation of the solar
system and contain information about early dynamical phases. As representatives
of the population of planetesimals that formed the terrestrial planets,
they also provide constraints on planet formation and composition models. 
To gain insight into these topics, however, we also need to fully understand
the evolutionary processes that asteroids have undergone such as collisions,
differentiation, or space weathering, and what imprints these have left on
their physical properties; for example, their size frequency distributions,
shapes, rotational states, and surface physical properties 
\citep[for a review, see][]{Michel2015}. 

Asteroid diameters can be estimated by fitting models of their thermal emission
to observations at thermal infrared wavelengths, at which said emission peaks as
a consequence of their typical surface temperatures. This technique, called
radiometry, has been the major contributor of asteroid diameters since the
1980s, when the Infrared Astronomical Satellite (IRAS) allowed two
thousand asteroid diameters to be determined \citep{Tedesco1986,Tedesco2002}. 
After a modest addition of observations of approximately 150 more asteroids by
the Midcourse Space Experiment (MSX) in the 1990s \citep{Tedesco2002b}, there
has been two major space-based all-sky surveys at thermal infrared wavelengths
in the last ten years: AKARI \citep{Murakami2007} and WISE/NEOWISE
\citep{Wright2010,Mainzer2011}. 

\citet{Usui2011}, henceforth U11, used the 9- and 18-$\mu$m AKARI all-sky fluxes
 to produce the Asteroid catalogue using AKARI (AcuA), comprising 5000 
asteroid diameters and visible geometric albedos based on an empirical 
calibration of the standard thermal model (STM) 
\citep[see e.g.][]{Lebofsky1986}. After the release of the AcuA, the AKARI flux
catalogue was updated with serendipitous asteroid detections among the IRC 
slow-scan observations \citep{Takita2012}, which were used by 
\citet{Hasegawa2013}, H13, to add 88 completely new or updated diameters and 
visible geometric albedos. These works completed the catalogue of 
main-belt asteroids (MBA) down to diameters of $\sim$20~km 
\citep{Usui2013,Usui2014}. 

If AKARI doubled the IRAS catalogue in terms of numbers, WISE/NEOWISE increased
the number by two orders of magnitude, providing observations of 
hundreds of near-Earth asteroids (NEAs) \citep{Mainzer2011}, more than a 
hundred thousand MBAs \citep{Masiero2011}, and thousands 
of outer MBAs and Jupiter Trojans \citep{Grav2012}. Still, as discussed by 
\citet{Usui2014}, the AKARI and IRAS catalogues constitute important 
complements to NEOWISE at the large-size end of the distribution by 
compensating for the absence of a small but crucial number of large asteroids.
In addition, the WISE detectors saturated with the largest targets, and
although partial saturation corrections have been applied successfully
\citep{Mainzer2011b} to estimate diameters using a thermal model, they may not
be optimal for more sophisticated thermo-physical models. Furthermore, the
available catalogues also complement each other in that they provide coverage
of different parts of the surfaces of many asteroids at different epochs and,
especially those with more eccentric orbits, at different observation
geometries. In this sense, AKARI is the longest-lasting fully-cryogenic survey
with a duration of 18 months, whereas WISE/NEOWISE (WISE Cryogenic Survey) and
IRAS spanned 8 and 9 months, respectively. 

In this work, our aims are to provide the AKARI IRC flux catalogue to the
community and to discuss several scientific applications to asteroid studies.
As an example, we recompute asteroid diameters with the near-Earth asteroid
thermal model \citep[NEATM;][]{Harris1998} with an upgraded version of the
implementation used in \citet{Ali-Lagoa2017}. The NEATM has become the
  standard method in the field, so this new catalogue can facilitate comparison
  with others based on space IR surveys, like the one by \citet{Ryan2010} from
  IRAS/MSX data, or \citet{Mainzer2016} based on NEOWISE. Our visible
geometric albedos are computed from our new diameters and the IAU 
  $H$-$G_{1}$-$G_{2}$ system \citep{Muinonen2010}. More specifically, we took the
  $H$-$G_{12}$ values provided by \citet{Oszkiewicz2011}; this alternative is
  better suited for cases where the optical data coverage is not sufficient to
  derive $G_1$ and $G_2$.

The structure of the article is as follows. In Sect.~\ref{sec:production} we
provide relevant information and references about the production of the flux
catalogue (Sect.~\ref{sec:fluxes}), our thermal model approach and its
validation (Sects.~\ref{sec:model} and \ref{sec:validation}), and we show a
sample of our new diameters catalogue (Sect.~\ref{sec:sample}). In
Sect.~\ref{sec:discussion} we compare our catalogue with the previous version
and with WISE/NEOWISE diameters and albedos (Sects.~\ref{sec:comparisonU11} and
\ref{sec:NEOWISE}), and we discuss several particular cases to illustrate the
advantages and disadvantages of our approach in the context of the beaming
parameter (Sect.~\ref{sec:eta}). Finally, in Sect.~\ref{sec:outlook}, we
conclude with some further expected scientific applications of the AKARI IRC
All-sky Survey asteroid flux catalogue.

\section{Production of the catalogue}\label{sec:production}
\subsection{AKARI Infrared Camera All-sky Survey asteroid fluxes}
\label{sec:fluxes}

The AKARI IRC All-sky Survey asteroid flux catalogue has been publicly
available in the JAXA website since October 2016\footnote{\scriptsize http://www.ir.isas.jaxa.jp/AKARI/Archive/Catalogues/Asteroid\_Flux\_V1/}.
Details about the mission concept \citep{Murakami2007}, the Infrared Camera
(IRC) \citep{Onaka2007}, point-source detection and calibration
\citep{Ishihara2010}, and the production of the asteroid catalogue in
particular were provided by \citet{Usui2011}. Likewise, \citet{Hasegawa2013}
explained the procedures followed to obtain the 89 slow-scan detections of 88
MBAs. In total, we used 20773 observations. 

The IRC asteroid flux catalogue includes data in bands \textit{S9W} and
\textit{L18W}, ranging from 6.7 to 11.6~$\mu$m and from 13.9 to 25.6~$\mu$m,
respectively. In addition to the measured fluxes, the catalogue contains the
asteroid number, name, and/or provisional designation, the epoch, sky
coordinates, and the geometry of observation, that is, the heliocentric distance
($r$), geocentric distance ($\Delta$), phase angle ($\alpha$), the angle
subtended by the Sun and the observer as seen from the asteroid, and the solar
elongation which, by the spacecraft's design, is always within 1$^\circ$ from
90$^\circ$ (this configuration is called ``quadrature'').

It is necessary to colour-correct the reported fluxes because the spectral
energy distributions (SEDs) of asteroids differ significantly over the widths
of the two bandpasses from those of the K- and M-type giant stars used as
calibrators. 
Colour-corrected fluxes and corresponding error bars are also given in the
catalogue. They are based on a third-order polynomial fit \citep[see Table 2 in
][]{Usui2011} to the colour corrections computed for a 10\%-albedo asteroid as
a function of heliocentric distance. 
While the temperature at a given heliocentric distance can differ by a few K
for very low- or high-albedo asteroids, we estimated that the corresponding
colour corrections do not vary significantly to introduce a strong bias,
especially for L18W data. For very-low-albedo Jupiter Trojans, however, this
approximation could result in colour corrections for band S9W being
underestimated by up to 4\%, so it may be advisable to recompute them for the
purposes of thermo-physical modelling. 
The filter response functions \citep{Onaka2007} are available at the JAXA
website\footnote{
  \scriptsize http://www.ir.isas.jaxa.jp/AKARI/Observation/support/IRC/RSRF/}.
For our purposes here this is not so determinant, as fitting the
beaming parameter can easily compensate for this small effect.

\subsection{Thermal modelling}\label{sec:model}

We used the near-Earth asteroid thermal model of \citet{Harris1998} (NEATM) as
implemented in \citet{Ali-Lagoa2017}, where the method is described in detail.
In summary, the NEATM  approximates the asteroid as a non-rotating sphere with
idealised surface properties: it behaves as a grey body with constant
emissivity of 0.9 and it does not conduct heat towards the subsurface, that is, 
each surface element reaches thermal equilibrium instantaneously so there is no
thermal inertia \citep[see][and references therein]{Harris2002,Delbo2015}. To
compensate for all simplifying assumptions and better fit thermal infrared
data, the NEATM uses the infrared beaming parameter $\eta$ as a free parameter.
It was empirically introduced in the framework of the standard thermal model to
account for the effects of surface roughness and the tendency of rough surfaces
to ``beam'' their thermal emission in the sunward direction
\citep{Lebofsky1986}. 

\begin{figure}
  \begin{center}
    \includegraphics[width=\hsize]{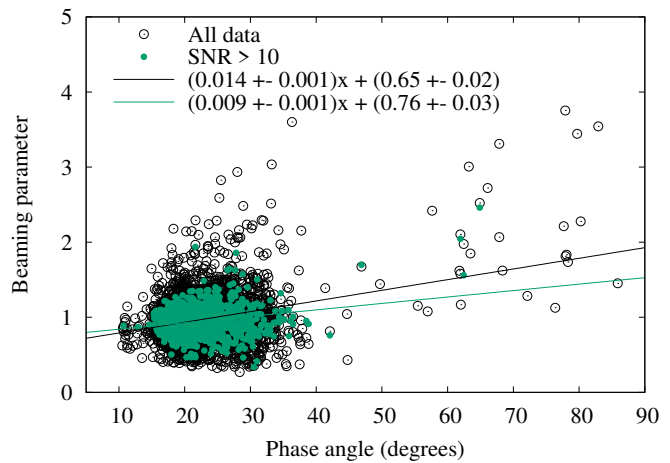}
    \caption{Best-fitting beaming parameter values as a function of phase angle
      and corresponding best-fitting straight lines. Black circles: fits based
      on the complete catalogue ($\approx$3000 fits). Green circles: fits based
      on two or more data points in both AKARI bands with S/N$>$10
      ($\approx$1000 fits). \label{fig:eta_vs_alpha}}
  \end{center}
\end{figure}

The NEATM was conceived to obtain more accurate diameters of near-Earth
asteroids (NEAs) as the empirical phase factor calibrated from large main-belt 
asteroid observations was not appropriate for the typically high phase 
angles at which NEAs are observed. Since the fitted beaming parameter values
correlate with the phase angle of the observation, $\alpha$
\citep{Delbo2003,Wolters2009}, it is customary to use a default value
($\eta_d$) based on the average of particular asteroid populations whenever it
is not possible to fit it (two or more thermal bands are required to have the
necessary degree of freedom). An average value of 1.4 was found for NEAs
\citep{Mainzer2011}, 1.2 for Mars-crossing asteroids \citep{Ali-Lagoa2017}, 1.0
for MBAs \citep{Masiero2011}, and 0.77 for Hildas and Jupiter
Trojans \citep{Grav2012}. Here we considered our sample's statistics to choose
a default value of $\eta$ as a function of $\alpha$.
Figure~\ref{fig:eta_vs_alpha} shows all our $\eta$-values based on two or more
data points with S/N$>10$ in each band as a function of phase angle (green
circles) and the corresponding best-fitting straight line, namely
\begin{equation}
  \eta_d(\alpha)=(0.009\pm0.001)~\mathrm{deg}^{-1}\alpha +(0.76\pm 0.03).
  \label{eq:eta}
\end{equation} 
The Figure also illustrates how taking data with S/N$>10$ only and
requiring at least two measurements in each band removes the majority of
extreme values of beaming parameter outside physical limits (usually set at
$\sim$0.5 and $\pi$). But since the IRC survey obtained typically two
observations per sighting per object, which is not enough to gain good
rotational coverage, it is still possible to obtain some high/low
$\eta$-values when fitting data even with reasonable S/N ratios (e.g.
elongated objects observed when their visible projected area was
minimum/maximum). To mitigate this, we did not include any fit with
$\eta$-values outside the 5-$\sigma$ limits defined by the uncertainties of
the coefficients in Eq. 1. Instead, we reran our model with a default value
of $\eta$ taken from Eq.~\ref{eq:eta}.

Arguably, fitting a straight line to the
correlation between the NEATM beaming parameter and the phase angle of
observation (Fig.~1) might not be the optimal option as the fit is not
statistically robust. However, although the correlation has been reported and
discussed several times in the literature \citep[e.g.][]{Delbo2003,
  Wolters2009,Masiero2011,Harris2016}, it is still not well understood on
purely physical grounds, likely because the beaming parameter is not a
physical quantity. Thus, we consider it preferable to fit a straight line to
our sample rather than any other function that may provide a statistically
better fit. Furthermore, Eq. 1 reproduces diameters of objects in both
extremes of the phase angle range satisfactorily and leads to default values
of $\eta$ that are similar to those adopted for NEAs and Jupiter Trojans in
NEOWISE works. 

Regardless of whether $\eta$ was fitted or not, in the very few cases
when our fits lead to unrealistic values of visible geometric albedos (we chose
the limits to be $p_V>1.0$ or $p_V<0.025$), we reran the model increasing or 
decreasing the beaming parameter by 15\%. For instance, improving a fit
that leads to an unrealistically low albedo requires a lower beaming parameter
since this leads to a smaller diameter, which in turn would result in a higher
value of $p_V$ given the $p_V$-$D$-$H$ relation: 
\begin{equation}
  p_V=\left(\frac{D_0}{D}10^{-H/5}\right)^2,
  \label{eq:albedo}
\end{equation}
where $H$ is the absolute magnitude and $D_0=1329~\mathrm{km}$
\citep[e.g.][]{Pravec2007}. 
We note that readjusting the beaming parameter based on an unrealistic
value of albedo can be counterproductive if it is the $H$ value and not the
thermal IR flux that is actually producing such an albedo value. However,
given that we sometimes only have one IR flux to fit, we consider this approach
more robust. 

We used the $H$-$G_{12}$ values computed by 
\citet{Oszkiewicz2011}, from now on referred to as O11. \citet{Veres2015}
(V15) also produced a  catalogue of $H$-$G_{12}$ values based on Pan-STARRS
photometry and found systematic differences with respect to the Minor Planet
Center and O11 absolute magnitudes but not with respect to those obtained by
\citet{Pravec2012} based only on high-quality photometry. However, O11 and V15
have different strengths (see the discussion in V15), and a considerable
fraction of objects in our catalogue are not included in the V15 catalogue. At
any rate, it is possible to update the albedos in our catalogue with new $H$
values using Eq.~\ref{eq:albedo}, but making a detailed comparison is beyond
the scope of this work. 

\subsection{Validation} \label{sec:validation}

\begin{figure}
  \begin{center}
    \includegraphics[width=\hsize]{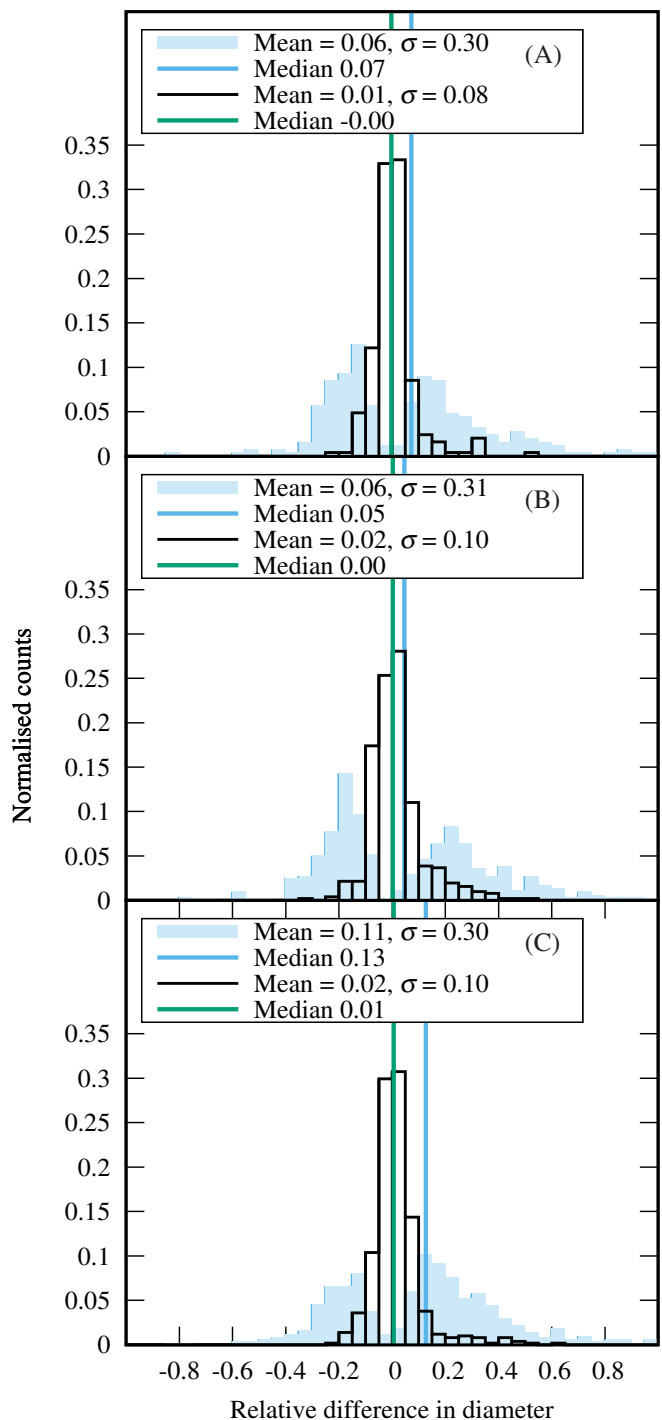}
    \put(-60,522){(A)}
    \put(-60,354){(B)}
    \put(-60,187){(C)}
    \caption{Maximum (blue) and minimum (empty boxes) relative differences 
      between radiometric sizes and occultation sizes, i.e. 
      $(D - x_{\mathrm{Occ}})/x_{\mathrm{Occ}}$, where $x_{\mathrm{Occ}}$ is either the 
      long axis, the minor axis, or the equivalent diameter of the ellipse 
      fitted to the occultation data. Panel A: Usui et al. (2011). Panel B: 
      Masiero et al. (2016). Panel C: this work.} \label{fig:eD_OCC}
  \end{center} 
\end{figure}

To validate our model, in this Section we compare the radiometric diameters of
U11/H13, NEOWISE \citep{Mainzer2016}, and this work, with diameters derived
from fits to stellar occultation chords compiled by \citet{Dunham2016}. We
computed the relative differences between each value of radiometric diameter
and $a$  and $D_{\mathrm{eq}}=\sqrt{ab}$, where $a$ and $b$ are the major and
minor axes of the best-fitting ellipse and $D_{\mathrm{eq}}$ is the diameter of
the circle with the same area. For each entry, we took the minimum and the
maximum differences and plotted them in Fig.~\ref{fig:eD_OCC}, which shows the
histograms we obtained for the diameters of U11H13 (A; it includes 184
objects), NEOWISE (B, 178 objects), and this work (C, 184 objects). The mean
and average values of the minimum relative differences (empty boxes) are all
close to zero and the standard deviations are of the order of 10\%, in
consistency with minimum NEATM uncertainty estimates of other previous works
\citep{Harris2006,Mainzer2011b}. 

The maximum relative differences are more widely spread ($\sigma\approx 30$\%)
and deviate somewhat more systematically from zero, with averages of $\sim 6$\%
for U11H13 and NEOWISE, and $\sim 10$\% for our diameters. This feature is not
unexpected since radiometric diameters derived from high-quality thermal IR
data are more likely to capture different cross-sections of irregular bodies as
they rotate and thus provide a reliable volume-equivalent diameter. On the
other hand, very accurate occultation chords are two-dimensional measures of
size and, albeit ``ground truth'' estimates, they are more likely to lead to
underestimates of the diameter and volume of very irregular bodies.
In addition, the fact that the AKARI IRC survey provided fewer observations per
sighting than WISE/NEOWISE could also explain why we obtain a higher average
maximum deviation for our sample, as it does not guarantee an average of large
and small cross-sections (see Sect.~\ref{sec:eta} for an example and further
discussion).

\subsection{The new diameter, albedo, and beaming parameter catalogue}
\label{sec:sample}

Table~\ref{tab:catalogue} shows a sample with a few lines of our new catalogue,
which contains a total of 8097 fits for 5198 different bodies. Some 
asteroids have more than one entry in our new catalogue because the beaming
parameter depends on the geometry of observation and requires that groups of
observations of the same object taken at widely different epochs be   
fitted separately. Following \citep{Mainzer2011b}, we set three days as the
maximum separation between consecutive data to be modelled together. 

The diameters, albedos, and beaming parameters are provided with other relevant
information, such as the $H$-$G_{12}$ values
\citep{Muinonen2010,Oszkiewicz2011}, the number of data modelled in each band,
the geometry of observation obtained from the Miriade ephemerides server
(http://vo.imcce.fr/webservices/miriade/), and the epoch. 
\begin{table*}
  \begin{center}
    \caption{Derived diameters ($D$), visible geometric albedos ($p_V$), and
      beaming parameters ($\eta$). The albedos were computed from the absolute
      asteroid magnitudes $H$ taken from \citet{Oszkiewicz2011}. Entries with
      ``True'' in the ``Fit'' column indicate that the beaming parameter was
      fitted; otherwise, the value was computed from Eq.~\ref{eq:eta}. On the
      other hand, the diameters were always fitted. Columns n$_{\mathrm{9}}$ and
      n$_{\mathrm{18}}$ refer to the number of observations fitted in the 9-$\mu$m
      and 18-$\mu$ bands, and $r$, $\Delta$, and $\alpha$ are the heliocentric
      and geocentric distances and phase angle, respectively. MJD is the
      modified Julian date of the last observation of the group. Typical
      minimum uncertainties are higher when the beaming parameter could not be
      fitted, as discussed in the main text. \label{tab:catalogue}}
    \begin{tabular}{ l  c  c  c  c  c  c  c  c c c c c}
    \hline\hline                 
    Object &  $H$ & $G_{12}$ & $D$ (km) & $p_V$  & $\eta$ & Fit    & n$_\mathrm{9}$ & n$_\mathrm{18}$ & $r$ (au) & $\Delta$ (au) & $\alpha$ (degree) & MJD \\
    \hline
    00001  &   3.43 &   0.88  &  1082.6 & 0.064 &  1.05 & True  & 2 &  1 & 2.95137 & 2.77621 & 20.01867 & 53868.7602 \\ 
    00001  &   3.43 &   0.88  &   929.9 & 0.087 & -0.94 & False & 0 &  2 & 2.98621 & 2.82770 & 19.37133 & 54048.8055 \\ 
    00001  &   3.43 &   0.88  &  1029.9 & 0.071 & -0.96 & False & 1 &  1 & 2.88090 & 2.69998 & 20.59184 & 54324.6499 \\ 
    00002  &   4.22 &   0.68  &  556.99 & 0.117 &  1.00 & True  & 3 &  3 & 3.38096 & 3.23953 & 17.24352 & 54006.1010 \\ 
    00002  &   4.22 &   0.68  &  535.46 & 0.126 &  0.99 & True  & 3 &  3 & 3.34383 & 3.18900 & 17.66837 & 54259.5766 \\ 
    00003  &   5.19 &   0.18  &  268.28 & 0.206 &  1.14 & True  & 3 &  2 & 2.96530 & 2.79400 & 19.37168 & 54116.1894 \\ 
    00003  &   5.19 &   0.18  &  247.44 & 0.242 &  1.08 & True  & 1 &  2 & 3.25764 & 3.10484 & 18.18754 & 54288.8980 \\ 
    00004  &   2.99 &   0.36  &  562.61 & 0.355 &  1.10 & True  & 2 &  3 & 2.18592 & 1.95199 & 26.91148 & 54154.8810 \\ 
    00005  &   6.84 &   0.30  &  105.40 & 0.292 &  0.83 & True  & 2 &  2 & 2.80221 & 2.61512 & 21.24749 & 53942.5128 \\ 
    00005  &   6.84 &   0.30  &  104.53 & 0.297 &  0.84 & True  & 2 &  1 & 2.45474 & 2.25927 & 23.63201 & 54119.4646 \\ 
    \end{tabular}
  \end{center}
\end{table*}

As in previous works, minimum error bars in $D$ are 10\% (20\% in $p_V$)
if the beaming parameter was fitted, and 20\% (40\% in $p_V$) otherwise.
Diameter fits to single data points are provided for completeness but should
be taken with caution.

\section{Discussion}\label{sec:discussion}
\subsection{Comparison with U11/H13}\label{sec:comparisonU11}

\begin{figure}
  \begin{center}
    \includegraphics[width=\hsize]{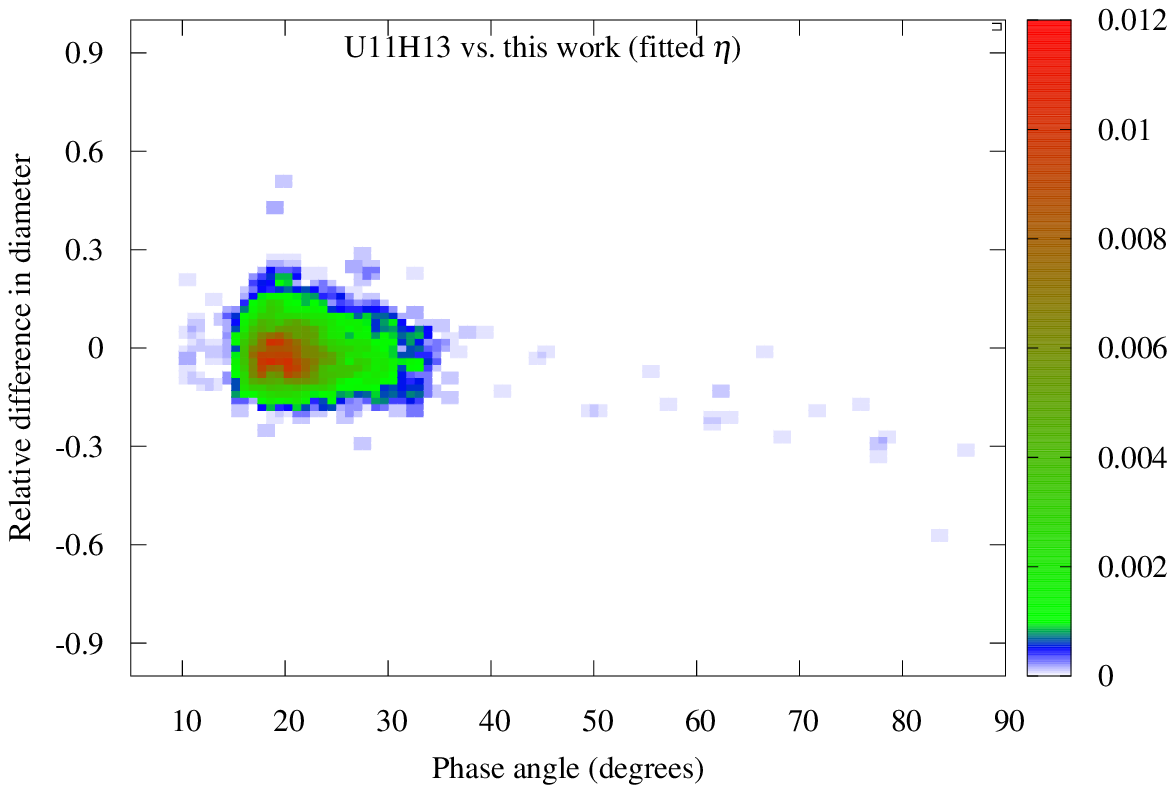}

    \vspace{-1.69cm}
    \includegraphics[width=\hsize]{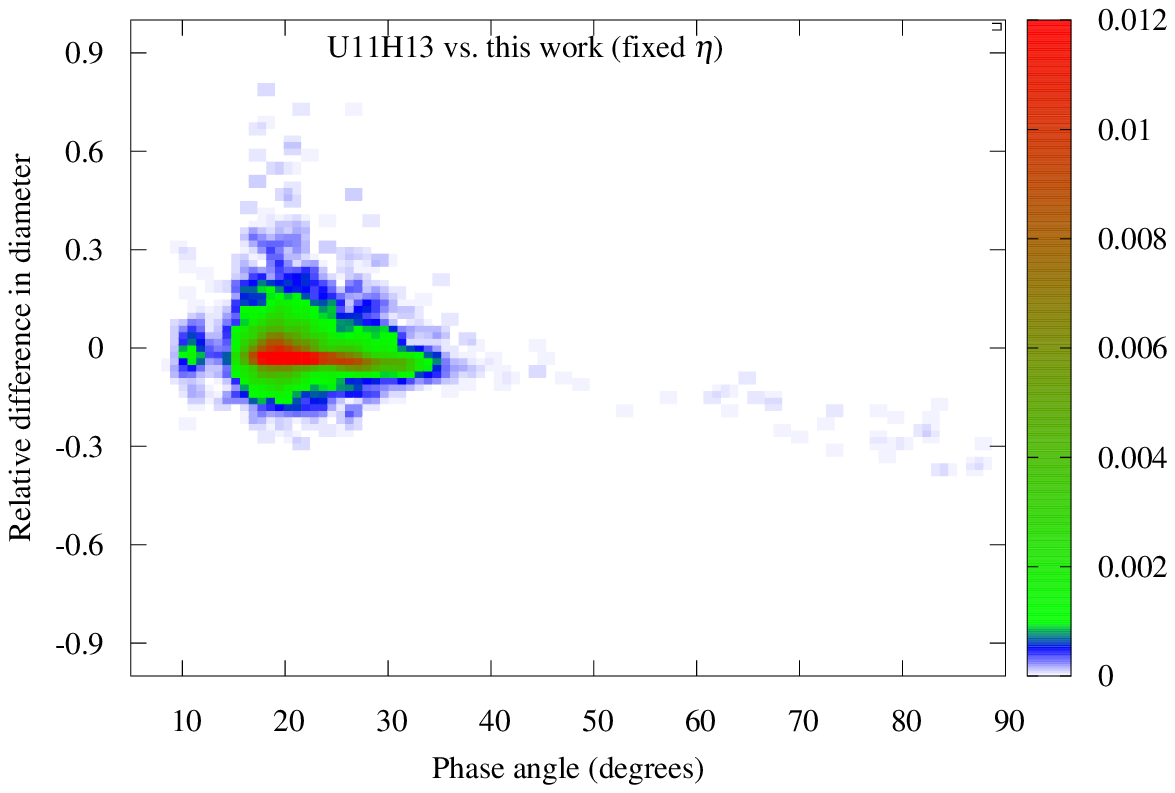}

    \caption{Relative differences between the diameters of U11/H13 and ours as
      a function of phase angle. The colour map indicates the number density of
      points in the box normalised to the total number of points in the plot.
      Top: cases for which $\eta$ could be fitted. Bottom: $\eta_d$ was taken
      as default. }\label{fig:comparison}
  \end{center} 
\end{figure}

In Sect.~\ref{sec:validation} we showed that we obtain statistically
equivalent diameters when we compare with occultation-based cross sections.
But the differences between our approach and that of U11/H13 become
systematically larger for objects observed at high phase angles, that is, NEAs
and some Mars-crossing asteroids. This is illustrated in
Fig.~\ref{fig:comparison}, where we plot the relative difference in diameter as
a function of phase angle for objects with fitted $\eta$ (top panel) and with
our calibrated default beaming parameter function, $\eta_d(\alpha)$ (lower
panel). The colour code, related to the density of points in the plot, shows
that a large fraction of cases at phase angles below 40$^\circ$ are well within
the expected $\pm$10\% relative differences, whereas the NEATM approach leads
to systematically larger diameters in the range $50^\circ < \alpha < 90^\circ$, in
creasing from 10\% to 20\%. An example is the potentially hazardous NEA (7341),
for which U11 reported $D =$~0.78~km and $p_V =$~0.62; instead, we found
$D =$~1.2~km and 1.1~km, and the corresponding $p_V =$~0.23 and 0.31 derived
for very similar $H$-values. Our albedos are more compatible with the spectral
classification of Sq found for this object \citep{Bus2002a,Bus2002b}.

Figure~\ref{fig:comparison} also shows a small component of objects at
intermediate phase angles for which we obtain significantly different
diameters. These are a consequence of a disadvantage in our approach: fitting
groups of data separately means that sometimes we are left with one or two
observations, often in a single band, which can lead to inaccurate diameters.
The reason not to average these diameters with other values in our catalogue
obtained for the same objects is, as pointed out in Sects.~\ref{sec:model} and
\ref{sec:eta}, to preserve potentially useful information about irregular
objects. In addition, it is not useful to average an inaccurate diameter with a
more reliable value. 

\begin{figure}
  \begin{center}
    \includegraphics[width=\hsize]{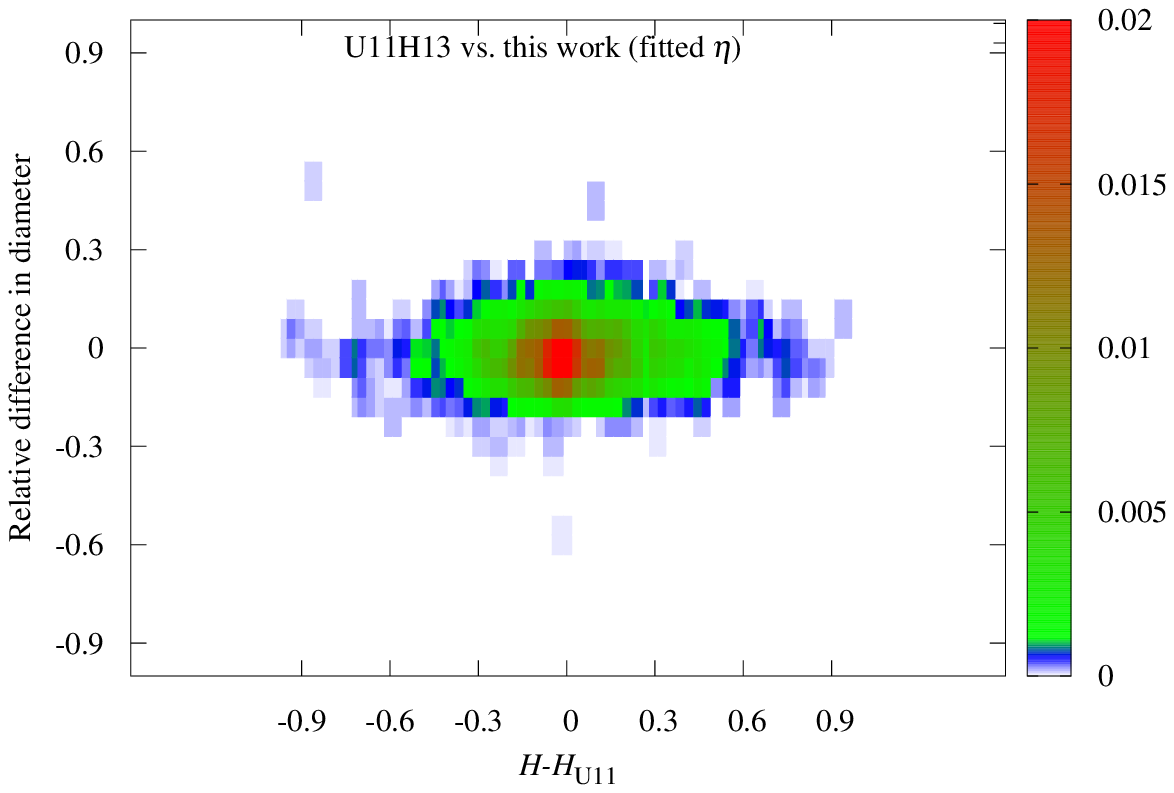}  
                                                                       
    \vspace{-1.70cm}                                                   
    \includegraphics[width=\hsize]{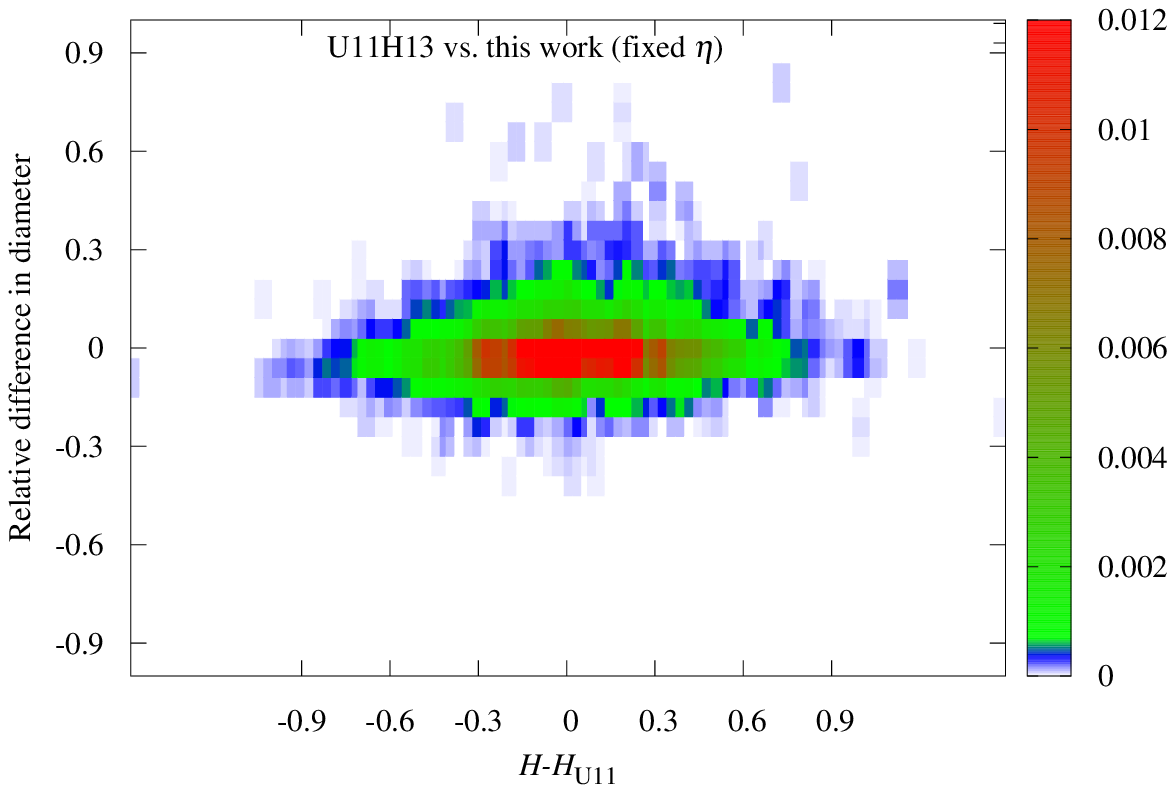}  
                                                                       
    \vspace{-1.70cm}                                                   
    \includegraphics[width=\hsize]{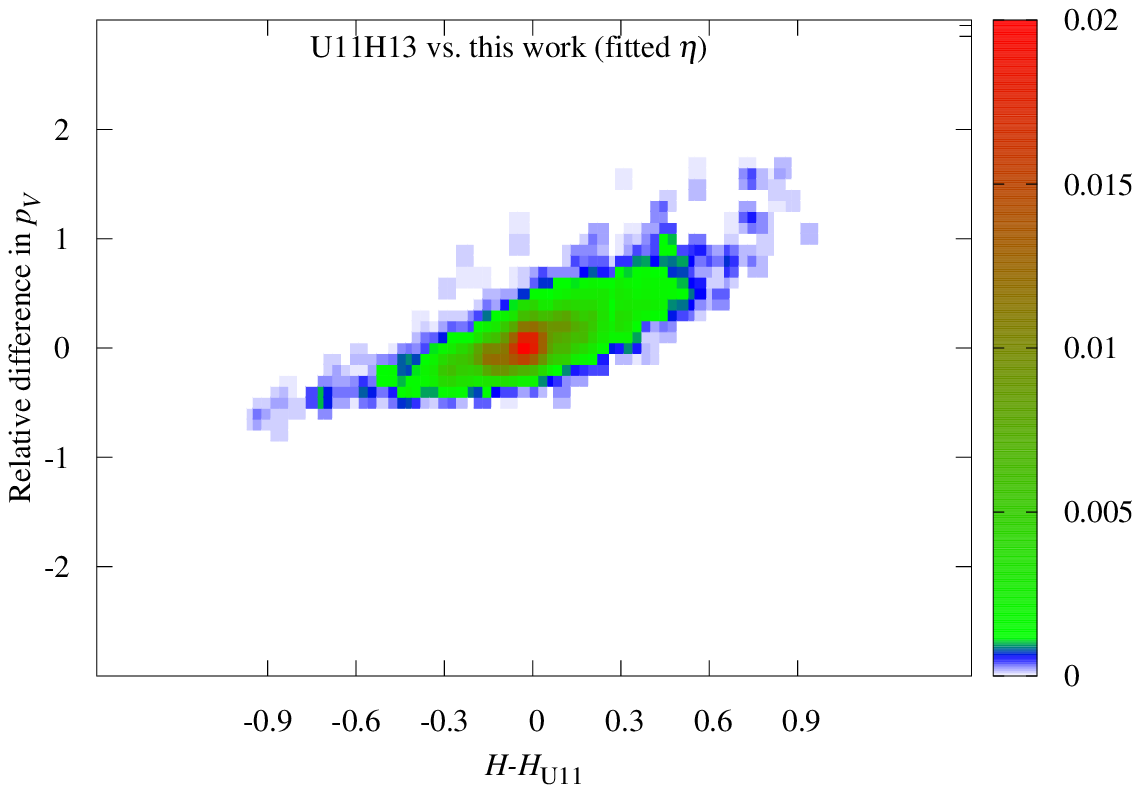} 
                                                                       
    \vspace{-1.70cm}                                                   
    \includegraphics[width=\hsize]{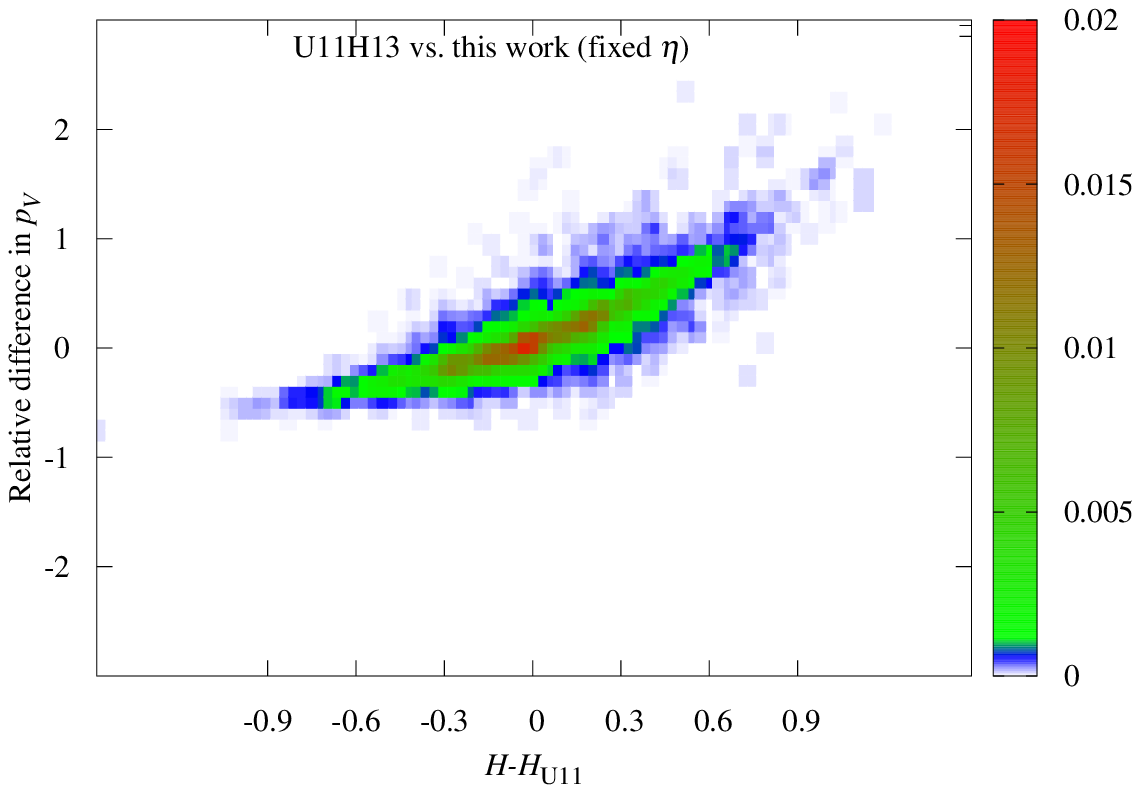} 

    \caption{Relative differences between the diameters of U11/H13 and ours
      (top panels) as a function of $H-H_{\mathrm{U11}}$. The colour map indicates
      the number density of points in the box normalised to the total number of
      points in the plot. The bottom panels show the same plot but for the
      visible geometric albedos.}\label{fig:eD_vs_DeltaH}
  \end{center} 
\end{figure}

From Eq.~\ref{eq:albedo}, the differences in albedo are a factor of two of the
differences in diameter if the $H$ value does not change. But many absolute
magnitudes have indeed been updated since U11 and H13, some by up to one
magnitude. Because the temperatures are proportional to $(1-A)^{\frac{1}{4}}$ and
the asteroids' Bond albedos $A$ are typically $< 0.4$, the impact of the
$H$-values on our diameters is small (see Sect.~3 in \citealt{Ali-Lagoa2017}).
We show this in Fig.~\ref{fig:eD_vs_DeltaH}, where the first two panels show
how the relative differences between the diameters of U11/H13 and ours are not
strongly correlated to the differences in $H$ values, even when we are not able
to fit $\eta$. The red and green areas, representing the large majority of our
sample, lie within the -0.30 to +0.30 bounds of the expected 2-$\sigma$
uncertainties. 
This is not the case for the visible geometric albedos, however. The computed
$p_V$ is a non-linear function of $H$, so we have an asymmetric trend in the
lower panels of the Figure (we highlight the change in scale of the $y$-axis).
For a small percentage of the sample the albedos have halved and some have even
decreased by a factor of 3. This resulted in an improvement in those cases when
U11/H13 reported unrealistically high values of $p_V$, such as (840), (1600),
or (3873), for example. On the other hand, we also obtained some
unrealistically low $p_V$-values ($< 0.025$). These were often based on fits to
a single data point, so we iteratively reran our model increasing the
$\eta_d$-value by 10\% until a more physical value of albedo was obtained. 
\begin{figure}
  \begin{center}
    \includegraphics[width=\hsize]{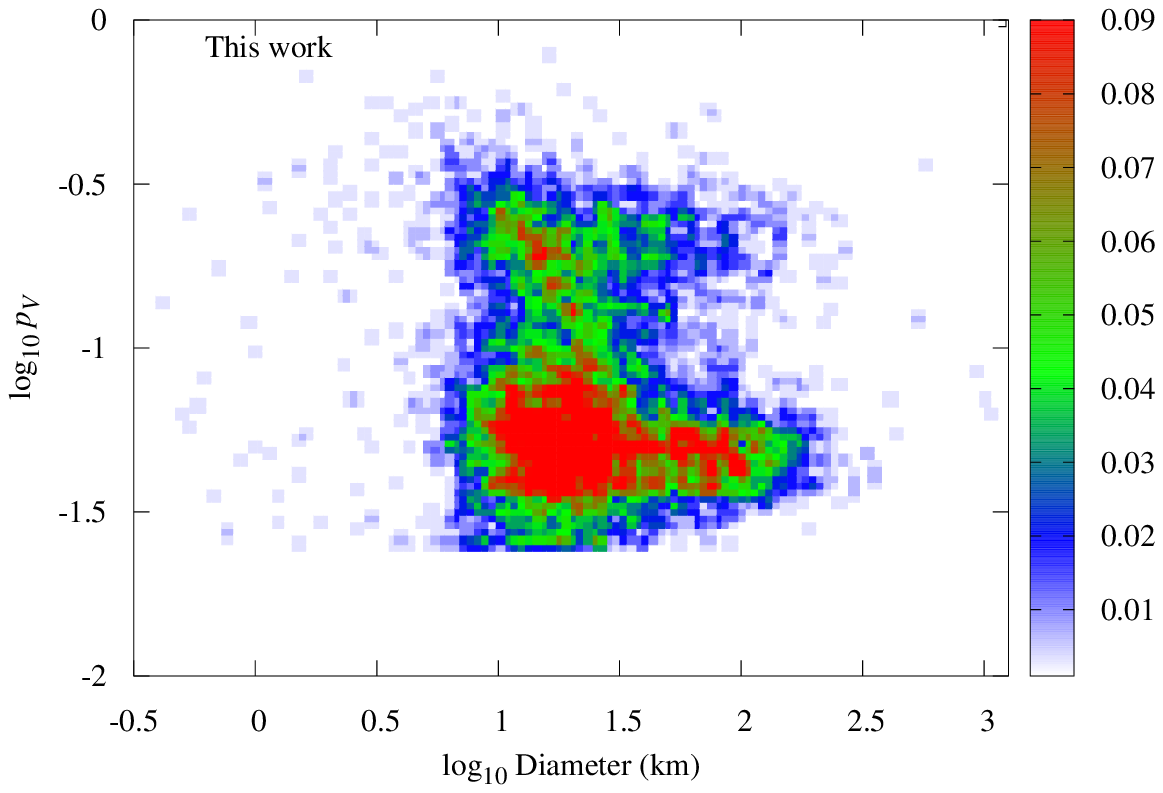} 
    
    \vspace{-0.5cm}                                                 
    \includegraphics[width=\hsize]{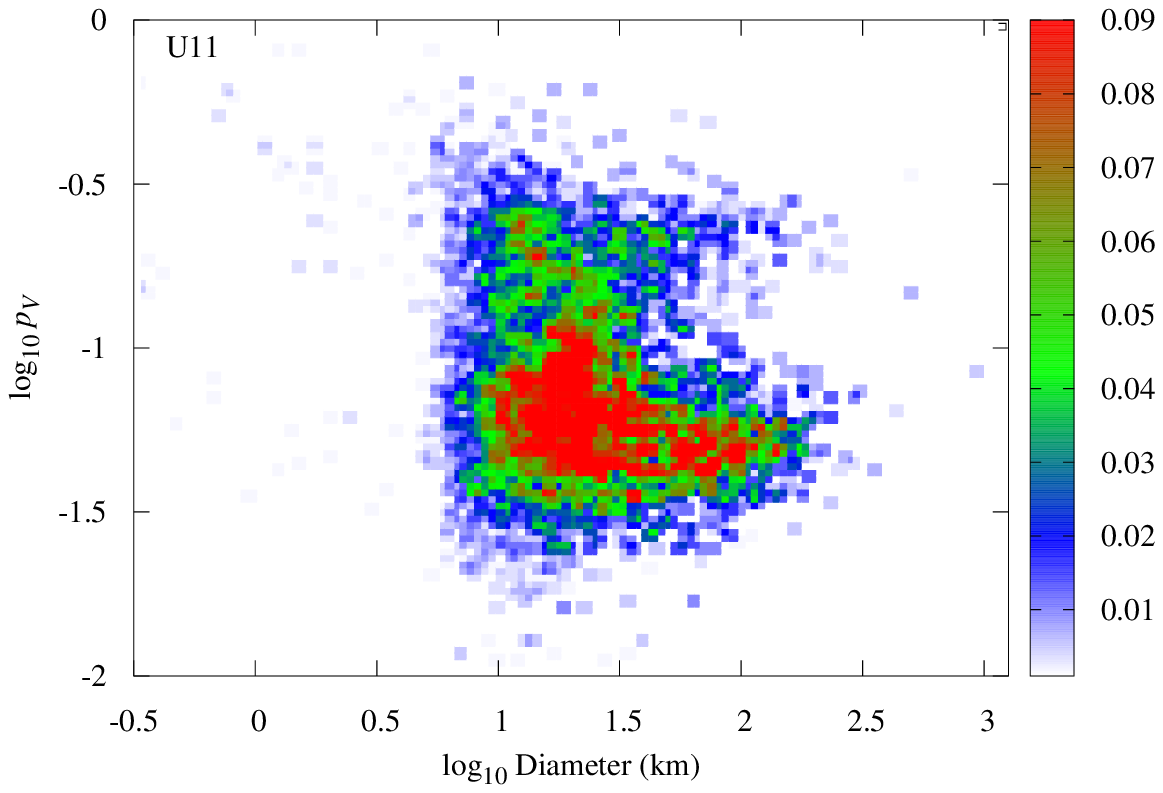}      
    
    \vspace{-0.5cm}                                                 
    \includegraphics[width=\hsize]{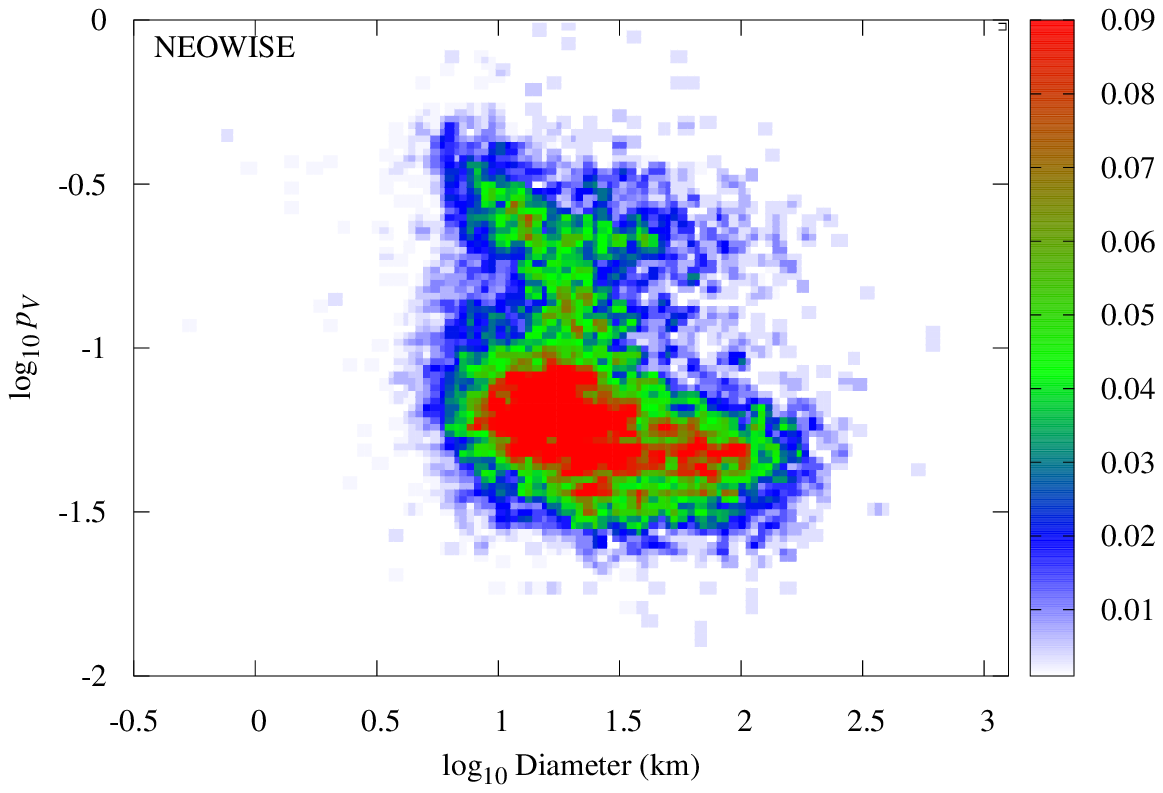}  
    
    \caption{Visible geometric albedos versus diameters derived here from the
      AKARI IRC All-sky survey fluxes (top panel), those of U11H13 (middle
      panel), and those computed for all our objects that were reported in the
      tables by \citet{Mainzer2016}. The colour code is proportional to the
      number of points in each bin normalised to the total number of points,
      which is different in each case. }\label{fig:pV_vs_D}
  \end{center} 
\end{figure}

\subsection{Intercomparison with WISE/NEOWISE diameters and albedos}
\label{sec:NEOWISE}

Figure~\ref{fig:pV_vs_D} shows density maps for plots of albedo versus diameter
for the whole sample computed here, the values of U11H13, and NEOWISE. We find
the two major albedo groups expected for the compositionally heterogeneous
asteroid population and very similar $p_V$-versus-$D$ plots when compared to U11
and NEOWISE \citep[for a comparative study of MBA albedo
  statistics based on AKARI and WISE/NEOWISE thermal models see][]{Usui2014}.
We find two differences worth mentioning. The intermediate albedo cluster
($0.10\,<p_V<\,0.20$) likely comprising X-complex asteroids
\citep{DeMeo2009,Mainzer2011e} seems to be more separated from the low-albedo
cloud in both our sample and the NEOWISE catalogue, whereas the U11H13
low-albedo cloud extends a little towards $p_V>0.10$. This could be partly
related to the update of many $H$ values for asteroids in the 10-30 km range
after the U11 work was completed. The second difference is that the NEOWISE
clusters stretch towards higher $p_V$ at lower sizes, which results in a
negative slope that is not apparent in the AKARI catalogues. This could be
partly due to the contribution of diameters derived from data taken during
non-cryogenic phases of the mission and the default values of beaming parameter
assumed, namely 1.0 \citep{Masiero2012} and 0.95 \citep{Nugent2015,Nugent2016}.
Those values are more representative of medium-sized and large objects
(10~km\,$< D <$\,100~km), whereas smaller bodies tend to require higher beaming
parameters
\citep[all other variables of the model being equal; see the discussions by][]{Ali-Lagoa2016,Ali-Lagoa2017}. 
Because underestimating $\eta$ results in a smaller diameter, it leads to a
higher albedo.

\subsection{Infrared beaming parameters}\label{sec:eta}

\begin{table*}
  \begin{center}
    \caption{Sizes of several asteroids based on radiometry from this work,
      U11/H13, and NEOWISE \citep[$D_{\mathrm{W}}$;][]{Mainzer2016}, and those
      determined from ellipsoids fitted to stellar occultation chords
      \citep[compiled by][]{Dunham2016}, and based on adaptive optics
      \citep[AO;][]{Hanus2017AO,Hanus2017}. Equivalent diameters from
      occultation and AO fits are also shown. The numbers in bold face indicate
      that the values of the beaming parameter for those fits were fixed to
      a default value. If several radiometric values are available, we have
      kept them in the same column, separated by commas. 
    }\label{tab:comparison}
    \begin{tabular}{ l  c  c  c  c  c  c  c }      
      \hline
      Asteroid & $D$ (km)    & $D_{\mathrm{U11}}$ (km) & $D_{\mathrm{W}}$  (km) & Occ. $a\times b$ (km) & $D_{\mathrm{Occ,eq}}$ (km) & AO $a\times b\times c$ (km)  & $D_{\mathrm{AO,eq}}$ (km) \\
      \hline
      \hline
      & & & & & & & \\
      (9)      & \textbf{223},~160 & 166           & 183,~184             & 176$\times$161  &  168                  & -                         & 168               \\
      &             &                     &                      & 200$\times$137  &  166                  &                           &                   \\
      &             &                     &                      & 182$\times$153  &  167                  &                           &                   \\
      & & & & & & & \\
      (10)     & 454,~403    & 428                 & \textbf{533}         & 464$\times$383  & 422                   & -                         & 412               \\
      & & & & & & & \\
      (16)     & \textbf{200},~236 & 207           & \textbf{288}         & 235$\times$230  & 233                   & -                         & 225               \\
      & & & & & & & \\
      (130)    & 212,~212    & 183                 & 181,\textbf{~162,~159} & 255$\times$154 &  199                 & 262$\times$205$\times$164 & 206               \\
      & & & & & & & \\
      (303)    & 107,~89     & 99,105               & 104,~\textbf{125}    & 87$\times$110    &  98                   & -                         & -                 \\
      & & & & & & & \\
      (1263)   & 50,~44       & 51                 & \textbf{40.2,~37.56} & 53.9$\times$36.2 & 44                    & -                         & -         \\                
      & & & & & & & \\
      \hline
    \end{tabular}
  \end{center}
\end{table*}

Table~\ref{tab:comparison} includes a sample of diameter fits that we obtained
for several objects compared to values taken from U11, the NEOWISE catalogue,
occultations, or adaptive optics. These examples show how the different
diameter fits obtained for the same object based on separate groups of
observations (see Sect.~\ref{sec:model}) are still compatible with the
dimensions derived from fits to stellar occultation data. This means that
high-quality AKARI data may be useful to identify irregular objects observed
more than once if they were observed at widely different aspects; we have
not averaged the diameter fits for this reason. 

\begin{figure}
  \begin{center}
    \includegraphics[width=\hsize]{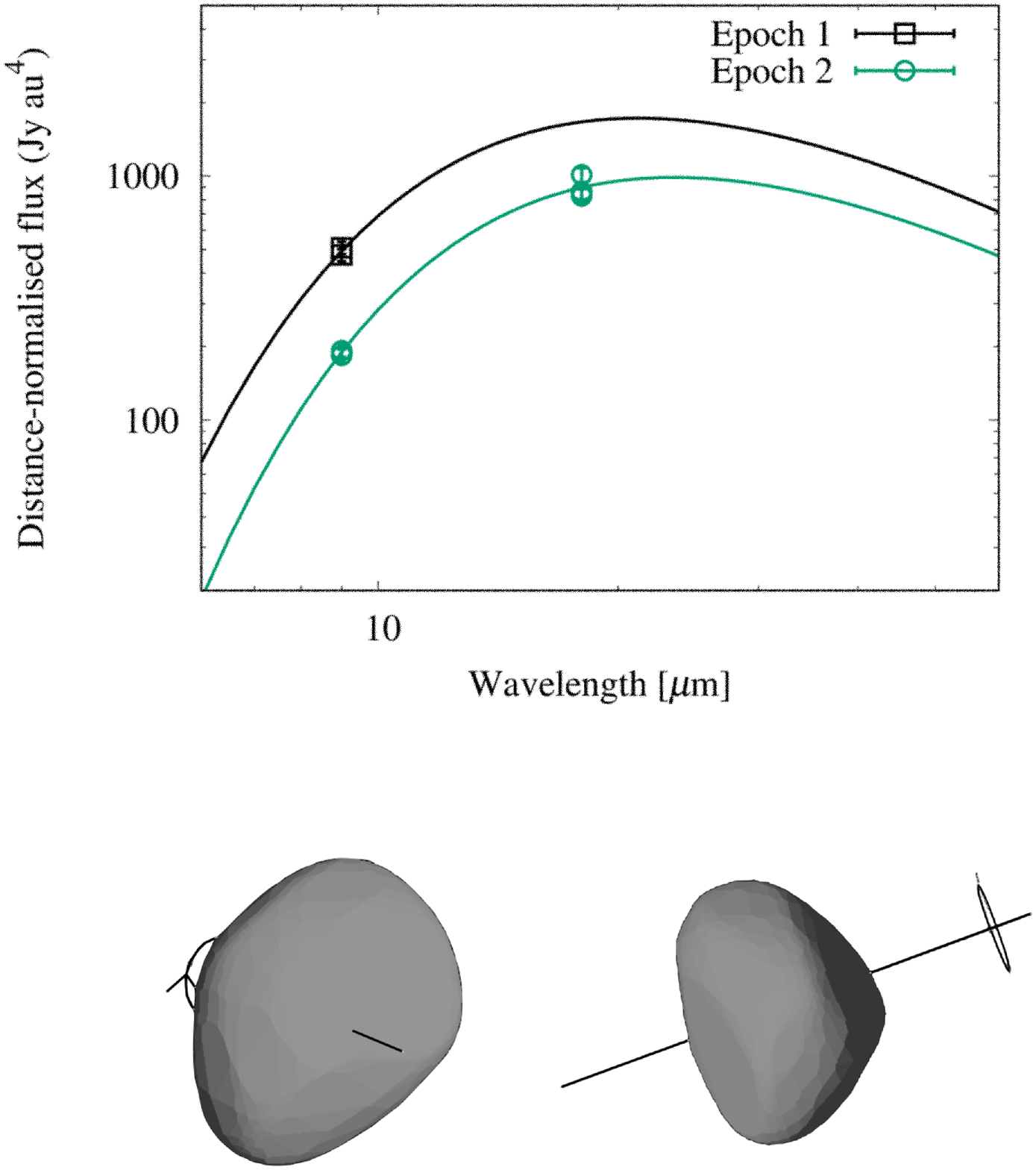}
    \put(-200,120){Epoch 1}
    \put(-90,120){Epoch 2}
    \vspace{-1.cm}
    \caption{Orientation of the convex model of (9) Metis
      \citep{Torppa2003,Durech2011} at the two epochs at which it was observed
      by AKARI. The almost pole-on view (Epoch 1) resulted in a larger observed
      projected area compared to the edge-on one (Epoch 2). This is
      contributing significantly to the ~20\% discrepancy between the two
      diameters.}\label{fig:metis}
  \end{center} 
\end{figure}

The case of (9) Metis is particularly interesting because it has also several
recorded occultation events with high-quality chords. AKARI observed Metis on
two occasions during the all-sky survey, so we derived two values for its
diameter that differ by $\approx$20\%. Compared to the high-quality occultation
fits\footnote{Quality codes 3 and 4.} by Dunham et al., it would be tempting to
reject our first diameter (223$\pm$40~km) over the second one (160$\pm$20~km).
However, we took the convex shape model of (9) Metis available in the DAMIT
database \citep{Durech2010,Durech2011} and rendered it\footnote{We used the
  ISAM service: http://isam.astro.amu.edu.pl \citep{Marciniak2012}.} as an
observer would have seen it at the different epochs. Figure~\ref{fig:metis}
shows that the cross-sectional area of the object at epoch 1, close to pole-on,
was larger than in the second epoch. We also checked that the three occultation
fits included in Table~\ref{tab:comparison} were taken at more edge-on
views\footnote{Incidentally, the only occultation event taken at a view closer
  to pole-on does present a larger cross section, but we did not include it in
  our table because it was assigned a quality code 2 by \citep{Dunham2016}}.

We also collected some other examples in Table~\ref{tab:comparison} that show
how choosing an inappropriate default value of $\eta$ may bias the radiometric
diameter. The diameters of (10) Hygiea and (16) Psyche given by 
\citet{Mainzer2016} were based on a default beaming parameter of 1.20 and 1.0,
respectively, whereas large main-belt asteroids' thermal data tend to be better
fitted by $\eta$-values in the range 0.7--0.9 (e.g. 
\citealt{Masiero2014,Ali-Lagoa2016}). Likely as a consequence, the
corresponding diameters are in both cases larger than those found from
occultations and adaptive optics. 

Conversely, because AKARI did not sample each object with as many observations
per sighting as WISE/NEOWISE (cf. from one to five data per band versus an
average of ten), some lower-quality AKARI fluxes occasionally lead to
unrealistic $\eta$ fits (as those apparent in Fig.~\ref{fig:eta_vs_alpha}) and
correspondingly inaccurate diameters and albedos. 
Thus, we do not expect this projected area effect to satisfactorily explain all
differences between radiometric diameters obtained from different epochs of
observations by AKARI or indeed any other survey. Given the current
visualisation tools that we have at our disposal, we cannot examine all
hundreds of cases individually, so we reiterate that these relative differences
should be taken as representative of the minimum error in diameter. In this
sense, because NEOWISE took more observations per sighting per object, it
samples the rotational phases better and is thus less prone to this problem
than AKARI. However, some degree of bias is still unavoidable, especially for
irregular objects that were observed only once (single apparition) at thermal
IR wavelengths. This effect will be even stronger on the values of the albedo
(Eq.~\ref{eq:albedo}), since the $H$ magnitudes are (usually) based on several
apparitions.

\section{Outlook and concluding remarks}\label{sec:outlook}

We provide the thermal infrared fluxes obtained for over 5000 asteroids by the
AKARI IRC All-sky survey, and an updated catalogue of diameters and albedos
based on the NEATM, which allowed us to fit the beaming parameter in numerous
cases. The visible geometric albedos were computed from the diameters and the
tabulated IAU $H$-$G_{12}$ parameters \citep{Oszkiewicz2011}. We validated our
approach by comparing our diameters with dimensions obtained from fits to
stellar occultation chords of over 100 asteroids, and we discussed the
usefulness of the catalogue by comparing its global properties against the AcuA
(U11,H13) and NEOWISE catalogues (Sects.~\ref{sec:comparisonU11} and
\ref{sec:NEOWISE}).

In Sect.~\ref{sec:eta} we focused on a few examples that illustrate how AKARI
IRC fluxes may provide three-dimensional information of elongated/irregular
asteroids based on the two or three groups of observations it recorded.
Although this is the main reason why we have not averaged our multiple
diameters, there are also some cases in which there were only one or two
measurements, often in one single band, so we argue that it is not advantageous
to average less accurate estimates with more reliable ones. By the same token,
we did not average with other diameters derived from WISE or IRAS because we
have not yet accumulated the minimum of ten or fifteen estimates per object
based on purely thermal data that would be required to obtain a statistically
robust result. To fulfill this aim, more all-sky surveys would be ideal
\citep[e.g.][]{Mainzer2016a}.

We emphasise two strengths of the AKARI catalogue versus the WISE/NEOWISE full
cryogenic survey, which is by far the largest source of asteroid thermal IR
data \citep[see the comparative study by][]{Usui2014}: the AKARI detectors did
not partially saturate for the largest targets, so AKARI is complete for
main-belt asteroids down to sizes of $\sim$20 km \citep{Usui2013,Usui2014}; the
all-sky survey lasted roughly 18 months, covering more than 95\% of the sky
twice (cf. $\sim$8 months of the WISE Cryogenic Survey, covering $\sim$20\%
twice). Specifically, $\sim$2000 asteroids in the catalogue were observed
twice, and $\sim 200$ three times. From these, we can anticipate further
scientific potential for more sophisticated thermo-physical modelling based on
the combination of AKARI, WISE/NEOWISE, IRAS, Spitzer, Herschel and other
sources \citep[e.g.][]{Mainzer2015}. For example, we are starting to gather
enough thermal IR observations before and after opposition for some key
objects, which would be especially helpful to constrain surface thermo-physical
properties whenever three-dimensional shape and spin axis orientation are
available. For those without shape information, it could be possible to
constrain the sense of rotation \citep{Mueller2002,MacLennan2013}. Also, some
bodies in highly eccentric orbits have been observed at different heliocentric
distances, which can open the possibility to study thermal inertia as a
function of temperature. Ultimately, more accurate diameters are crucial to
obtain accurate densities for the asteroids whose masses will be estimated
thanks to Gaia \citep{Tanga2012}, because the error in the diameter spreads
threefold onto the volume estimate and thus dominates the error in the density,
as pointed out by \citet{Carry2012}.

All these aspects are the focus of the ``Small Bodies: Near And Far'' (SBNAF),
a project funded by the EU to carry out a benchmark study on small-body
physical and thermal properties (three-dimensional shapes, rotational states,
thermal inertias, etc.) to assess our current models' state of the art and to
ensure a fruitful scientific exploitation of space- and ground-based data on
which these models are based \citep{Mueller2017cospar}. In a broader context, 
accurate knowledge of the physical properties of large main-belt asteroids is
extremely valuable for calibration purposes for ALMA and other
millimetre and sub-millimetre observatories \citep{Mueller2014}.

\begin{acknowledgements}
  The research leading to these results has received funding from the European
  Union’s Horizon 2020 Research and Innovation Programme, under Grant Agreement
  no 687378. This research is based on observations with AKARI, a JAXA project
  with the participation of ESA. The work of S.H. was supported by the JSPS
  KAKENHI Grant Numbers JP15K05277 and JP17K05636, and by the Hypervelocity
  Impact Facility (former facility name: the Space Plasma Laboratory), ISAS,
  JAXA.  This research has made use of data and/or services provided by the
  International Astronomical Union's Minor Planet Center.  This publication
  makes use of data products from NEOWISE, which is a project of the Jet
  Propulsion Laboratory/California Institute of Technology, funded by the
  Planetary Science Division of the National Aeronautics and Space
  Administration.  This research also made use of the NASA/ IPAC Infrared
  Science Archive, which is operated by the Jet Propulsion Laboratory,
  California Institute of Technology, under contract with the National
  Aeronautics and Space Administration. This research has made use of IMCCE's
  Miriade VO tool. The thermal model code was based on a development of
  M. Delbo's implementation \citep{Delbo2002}. 

\end{acknowledgements}

\bibliographystyle{aa}  
\bibliography{Asteroids} 

\end{document}